\documentclass[journal=jctcce,manuscript=article,layout=twocolumn]{achemso}
\usepackage[version=3]{mhchem}
\usepackage{amssymb}
\usepackage{color}
\usepackage{braket}
\usepackage{xspace}
\usepackage{cleveref}
\usepackage{graphicx}
\usepackage{subfig}
\usepackage{threeparttable}
\usepackage{textcomp}
\usepackage{booktabs}

\newcommand*{\mae}{$\Delta_{\mathrm{MAE}}$\xspace}
\newcommand*{\std}{$\Delta_{\mathrm{STD}}$\xspace}
\newcommand*{\maxe}{$\Delta_{\mathrm{MAX}}$\xspace}

\newcommand*{\cvs}{CVS-ODC-12\xspace}

\crefname{figure}{Figure}{Figures}      
\crefname{table}{Table}{Tables}         
\crefname{equation}{Eq.}{Eqs.}          
\crefname{section}{Section}{Sections}   
\crefname{subsection}{Section}{Sections}

 \author{Ruojing Peng}
 \affiliation{%
     Department of Chemistry and Biochemistry,
     The Ohio State University,
     Columbus, Ohio 43210, United States
 }
 \author{Andreas V.\ Copan}
 \affiliation{%
     Chemical Sciences and Engineering Division, 
     Argonne National Laboratory,
     Argonne, Illinois 60439, United States
 }
 \author{Alexander Yu.\ Sokolov}
 \email{sokolov.8@osu.edu}
 \affiliation{%
     Department of Chemistry and Biochemistry,
     The Ohio State University,
     Columbus, Ohio 43210, United States
 }

\begin{tocentry}
\includegraphics{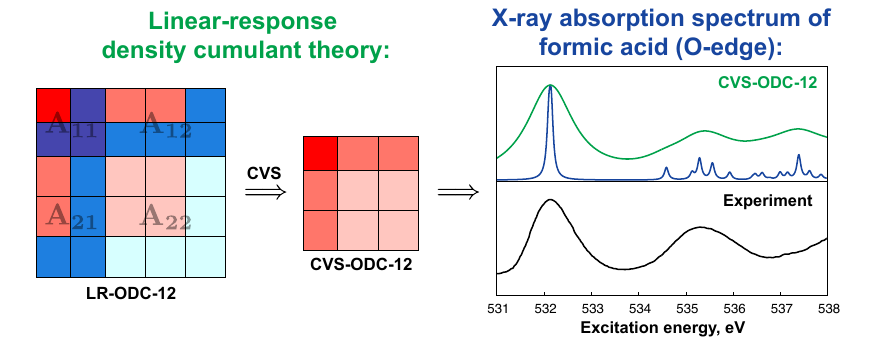}
\end{tocentry}

\title{Simulating X-ray Absorption Spectra With Linear-Response Density Cumulant Theory}

\begin{document}

\abstract{%
We present a new approach for simulating X-ray absorption spectra based on linear-response density cumulant theory (LR-DCT) [Copan, A.\@ V.; Sokolov, A.\@ Yu. {\it J. Chem. Theory Comput.}, {\bf 2018}, {\it 14}, 4097--4108]. Our new method combines the LR-ODC-12 formulation of LR-DCT with core-valence separation approximation (CVS) that allows to efficiently access high-energy core-excited states. We describe our computer implementation of the CVS-approximated LR-ODC-12 method (\cvs) and benchmark its performance by comparing simulated X-ray absorption spectra to those obtained from experiment for several small molecules. Our results demonstrate that the \cvs method shows a good agreement with experiment for relative spacings between transitions and their intensities, but the excitation energies are systematically overestimated. When comparing to results from excited-state coupled cluster methods with single and double excitations, the \cvs method shows a similar performance for intensities and peak separations, while coupled cluster spectra are less shifted, relative to experiment. An important advantage of \cvs is  that its excitation energies are computed by diagonalizing a Hermitian matrix, which enables efficient computation of transition intensities. 
}

\section{Introduction}
\label{sec:intro}

Near-edge X-ray absorption spectroscopy (NEXAS) is a powerful and versatile experimental technique for determining the geometric and electronic structure of a wide range of chemical systems. The NEXAS spectra probe excitations of core electrons into the low-lying unoccupied molecular orbitals. Due to the localized nature of core orbitals, these excitations are very sensitive to the local chemical environment, providing important information about molecular structure. Recent advances in experimental techniques for generating and detecting X-ray radiation have spurred the development of NEXAS and its applications in chemistry and biology.\cite{Contini:2001p7308,Wernet:2004p995,Freiwald:2004p413,Hahner:2006p1244,Plekan:2008p360,Hua:2010p13214,Guo:2011p181901} 

Theoretical simulations of X-ray absorption play a critical role in interpretation of the NEXAS spectra.\cite{Norman:2018p7208} However, computations of the core-level excitations are very challenging as they require simulating excited states selectively in the high-energy spectral region and a balanced treatment of electron correlation, orbital relaxation, and relativistic effects, often combined with large uncontracted basis sets. Many of the popular excited-state methods have been adopted for simulations of X-ray absorption spectra, including linear-response,\cite{Stener:2003p115,Besley:2010p12024,Liang:2011p3540,Zhang:2012p194306,Lestrange:2015p2994,Besley:2016p5018} real-time,\cite{Lopata:2012p3284,Fernando:2015p646,Bruner:2016p3741} and orthogonality-constrained\cite{Derricotte:2015p14360,Verma:2016p144} density functional theory, configuration interaction,\cite{Barth:1980p149,Butscher:1977p449,Butscher:1977p457,Asmuruf:2008p267,Roemelt:2013p3069,Roemelt:2013p204101} algebraic diagrammatic construction (ADC),\cite{Barth:1985p867,Koppel:1997p4415,Trofimov:2000p483,Plekan:2008p360,Wenzel:2014p1900,Wenzel:2015p214104} as well as linear-response (LR-) and equation-of-motion (EOM-) coupled cluster (CC) theories.\cite{Nooijen:1995p6735,Brabec:2012p171101,Sen:2013p2625,Coriani:2012p022507,Coriani:2012p1616,Kauczor:2013p211102,Fransson:2013p124311,List:2014p244107,Dutta:2014p3656,Peng:2015p4146,Coriani:2015p181103,Myhre:2016p2633,Nascimento:2017p2951,Lopez:2018} Among these approaches, CC methods have been shown to yield particularly accurate results for core excitation energies and intensities of small molecules. 

Several techniques to compute the NEXAS spectra within the framework of CC theory have been developed. These approaches usually incorporate up to single and double excitations in the description of electron correlation (CCSD), but employ different strategies to access high energies required to excite core electrons. For example, in the complex polarization propagator-based CC theory (CPP-CC),\cite{Coriani:2012p022507,Coriani:2012p1616,Kauczor:2013p211102,Fransson:2013p124311,List:2014p244107} core-level excitations are probed directly by computing the CC linear-response function over a grid of input frequencies in the X-ray region. In the energy-specific EOM-CC approach (ES-EOM-CC),\cite{Peng:2015p4146} the high-energy excitations are computed by using a frequency-dependent non-Hermitian eigensolver. In practice, both CPP-CC and ES-EOM-CC can only be applied to narrow spectral regions that need to be selected {\it a priori}. This problem is circumvented in the time-dependent,\cite{Nascimento:2017p2951} multilevel,\cite{Myhre:2016p2633} and core-valence-separated\cite{Coriani:2015p181103,Lopez:2018} EOM-CC methods that can be used to compute NEXAS spectra for broad spectral regions and a large number of electronic transitions.

In this work, we present an implementation of the recently developed linear-response density cumulant theory (LR-DCT)\cite{Copan:2018p4097} for simulating the NEXAS spectra of molecules. Although the origin of LR-DCT is in reduced density matrix theory,\cite{Colmenero:1993p979,Nakatsuji:1996p1039,Mazziotti:1998p4219,Kollmar:2006p084108,DePrince:2007p042501,DePrince:2016p164109} it has a close connection with LR-CC methods, such as linear-response formulations of linearized, unitary, and variational CC theory.\cite{Kutzelnigg:1991p349,Kutzelnigg:1998p65,VanVoorhis:2000p8873,Kutzelnigg:1982p3081,Bartlett:1989p133,Watts:1989p359,Szalay:1995p281} In our previous work, we have demonstrated that one of the LR-DCT methods (LR-ODC-12) provides very accurate description of electronic excitations in the UV/Vis spectral region.\cite{Copan:2018p4097} In particular, for a set of small molecules, LR-ODC-12 showed mean absolute errors in excitation energies of less than 0.1 eV, with a significant improvement over EOM-CC with single and double excitations (EOM-CCSD). We have also demonstrated that LR-ODC-12 provides accurate description of challenging doubly excited states in polyenes. 

Here, we test the accuracy of LR-ODC-12 for simulations of core-level excitations. To efficiently access the X-ray spectral region, our new LR-ODC-12 implementation employs the core-valence separation (CVS) technique,\cite{Cederbaum:1980p481,Cederbaum:1980p206} originally developed in the framework of ADC theory\cite{Barth:1985p867,Koppel:1997p4415,Trofimov:2000p483,Plekan:2008p360,Wenzel:2014p1900,Wenzel:2015p214104} and later extended to other methods.\cite{Stener:2003p115,Coriani:2015p181103,Lopez:2018} We test our new method (denoted as \cvs) against LR-ODC-12 to assess the accuracy of the CVS approximation and benchmark its results for a set of small molecules.

\section{Theory}
\label{sec:theory}

\subsection{Density Cumulant Theory (DCT)}

We start with a short overview of density cumulant theory (DCT).\cite{Kutzelnigg:2006p171101,Simmonett:2010p174122,Sokolov:2012p054105,Sokolov:2013p024107,Sokolov:2013p204110,Sokolov:2014p074111,Wang:2016p4833} 
The exact electronic energy of a stationary state $\ket{\Psi}$ can be expressed in the form
\begin{equation}
    \label{eq:energy-expression}
    E
    =
    \braket{\Psi|\hat{H}|\Psi}
    =
    \sum_{pq} h_p^q
    \gamma^p_q
    +
    \frac{1}{4}
    \sum_{pqrs} \overline{g}_{pq}^{rs}
    \gamma^{pq}_{rs}
\end{equation}
where the one- and antisymmetrized two-electron integrals (\(h_p^q\) and \(\overline{g}_{pq}^{rs}\)) are traced with the reduced one- and two-body density matrices ($\gamma^p_q$ and $\gamma^{pq}_{rs}$) over all spin-orbitals in a finite one-electron basis set. Starting with \cref{eq:energy-expression}, DCT expresses $\gamma^p_q$ and $\gamma^{pq}_{rs}$ in terms of the fully connected contribution to $\gamma^{pq}_{rs}$ called the two-body density cumulant ($\lambda^{pq}_{rs}$).\cite{Fulde:1991,Ziesche:1992p597,Kutzelnigg:1997p432,Mazziotti:1998p419,Mazziotti:1998p4219,Kutzelnigg:1999p2800,Ziesche:2000p33,Herbert:2007p261,Kong:2011p214109,Hanauer:2012p50} For $\gamma^{pq}_{rs}$, this is achieved using a cumulant expansion
\begin{equation}
    \label{eq:two-body-n-rep}
    \gamma^{pq}_{rs}
    =
    \gamma^p_r
    \gamma^q_s
    -
    \gamma^p_s
    \gamma^q_r
    +
    \lambda^{pq}_{rs}
\end{equation}
where the first two terms represent a disconnected antisymmetrized product of the one-body density matrices. To determine $\gamma^p_q$ from $\lambda^{pq}_{rs}$, the following non-linear relationship is used\cite{Sokolov:2013p024107}
\begin{equation}
    \label{eq:one-body-n-rep}
    \sum_{r} \gamma^p_r
    \gamma^r_q
    -
    \gamma^p_q
    =
    \sum_{r} \lambda^{pr}_{qr}
\end{equation}
where the r.h.s.\@ of \cref{eq:one-body-n-rep} contains a partial trace of density cumulant. \cref{eq:two-body-n-rep,eq:one-body-n-rep} are exact, so that substituting in the exact $\lambda^{pq}_{rs}$ yields the exact electronic densities and energy.

In practice, DCT computes the electronic energy by parametrizing and determining density cumulant directly, circumventing computation of the many-electron wavefunction. This is achieved by choosing a specific Ansatz for the wavefunction $\ket{\Psi}$ and expressing density cumulant as\cite{Sokolov:2014p074111}
\begin{equation}
    \label{eq:cumulant-parametrization}
    \lambda^{pq}_{rs}
    =
    \langle\Psi|
    a^{pq}_{rs}
    |\Psi\rangle_c
\end{equation}
where \(a^{pq}_{rs}\equiv a_p^\dagger a_q^\dagger a_s a_r\) is a two-body second-quantized operator and the subscript $c$ indicates that only fully connected terms are retained. The most commonly used parametrization of $\lambda^{pq}_{rs}$, denoted as ODC-12,\cite{Sokolov:2013p024107,Sokolov:2013p204110} consists of approximating \cref{eq:cumulant-parametrization} using a two-body unitary transformation of $\ket{\Psi}$ truncated at the second order in perturbation theory
\begin{align}                                                                                                           
    \label{eq:cumulant-parametrization_approx}                                                                          
    \lambda^{pq}_{rs}                                                                                                   
    &\approx                                                                                                            
    \langle\Phi| e^{-(\hat{T}^{}_2-\hat{T}^\dag_2)} a^{pq}_{rs} e^{\hat{T}^{}_2-\hat{T}^\dag_2} |\Phi\rangle_c \notag \\
    &\approx                                                                                                            
    \langle\Phi| a^{pq}_{rs}|\Phi\rangle_c                                                                              
    + \langle\Phi| [a^{pq}_{rs},\hat{T}^{}_2-\hat{T}^\dag_2]|\Phi\rangle_c \notag \\                                    
    &+ \frac{1}{2}\langle\Phi| [[a^{pq}_{rs},\hat{T}^{}_2-\hat{T}^\dag_2], \hat{T}^{}_2-\hat{T}^\dag_2]|\Phi\rangle_c   
\end{align}                                                                                                             
where \(\hat{T}_2\) is the double excitation operator with respect to the reference determinant \(\ket{\Phi}\), whose parameters are determined to make the electronic energy in \cref{eq:energy-expression} stationary. The ODC-12 energy is also made stationary with respect to the variation of molecular orbitals parametrized using the unitary singles operator \(e^{\hat{T}_1-\hat{T}_1^\dagger}\).\cite{Sokolov:2013p204110} The parameters of the \(\hat{T}_1\) and \(\hat{T}_2\) operators determined from the stationarity conditions are used to compute the ODC-12 energy.

\subsection{Linear-Response DCT (LR-DCT)}

In conventional DCT, electronic energy and molecular properties are determined for a single electronic state (usually, the ground state). To obtain access to excited states, we have recently combined DCT with linear-response theory that allows to compute excitation energies and transition properties for a large number of states simultaneously.\cite{Copan:2018p4097} In linear-response DCT (LR-DCT), we consider the behavior of an electronic system under a time-dependent perturbation \(\hat{V}f(t)\), which can be described using the time-dependent quasi-energy function\cite{Norman:2011p20519,Helgaker:2012p543} 
\begin{equation}
    \label{eq:quasi-energy-expression}
    Q(t)
    =
    \langle\Psi(t)|
        \hat{H}
        +
        \hat{V}f(t)
        -
        i\tfrac{\partial}{\partial t}
    |\Psi(t)\rangle
\end{equation}
Here, \(\ket{\Psi(t)}\) is the so-called ``phase-isolated'' wavefunction, which reduces to the usual time-independent wavefunction \(\ket{\Psi}\) in the stationary state limit. Importantly, for a periodic time-dependent perturbation, the quasi-energy averaged over a period of oscillation (\(\{Q(t)\}\)) is variational with respect to the exact time-dependent state.\cite{Helgaker:2012p543} Such periodicity implies that the amplitude \(f(t)\) can be written in a Fourier series
\begin{equation}
    f(t)
    =
    \sum_\omega f(\omega) e^{-i\omega t}
\end{equation}
where the sum includes positive and negative values for all frequencies such that \(f(t)\) is real-valued. 

To obtain information about excited states, \(\{Q(t)\}\) is made stationary with respect to all of the parameters that define the time-dependent wavefunction \(\ket{\Psi(t)}\).\cite{Kristensen:2009p044112} We refer interested readers to our previous publication\cite{Copan:2018p4097} for derivation of the LR-DCT equations and summarize only the main results here. The LR-DCT excitations energies are computed by solving the generalized eigenvalue problem
\begin{equation}
    \label{eq:excitation-energy}
    \mathbf{E}
    \mathbf{z}_k
    =
    \mathbf{M}
    \mathbf{z}_k
    \omega_k
\end{equation}
In \cref{eq:excitation-energy}, $\omega_k$ are the excitation energies, \(\mathbf{E}\) is the LR-DCT Hessian matrix that contains second derivatives of the electronic energy \(\{\langle\Psi(t)|\hat{H}|\Psi(t)\rangle\}\) with respect to parameters of the \(\hat{T}_1\) and \(\hat{T}_2\) operators, and \(\mathbf{M}\) is the metric matrix that originates from second derivatives of the time-derivative overlap \(\{\langle\Psi(t)|i\dot{\Psi}(t)\rangle\}\). Importantly, the LR-DCT Hessian matrix \(\mathbf{E}\) is Hermitian, which ensures that the excitation energies $\omega_k$ have real values, provided that the Hessian is positive semidefinite. The generalized eigenvectors $\mathbf{z}_k$ can be used to determine the oscillator strength for each transition
\begin{equation}
    \label{eq:transition-strength}
    f_{\mathrm{osc}} (\omega_k)
    =
    \frac{2}{3} \omega_k 
|\langle\Psi|\hat{V}|\Psi_k\rangle|^2
    =
    \frac{2}{3} \omega_k 
    \frac{%
        |\mathbf{z}_k^\dagger \mathbf{v}'|^2
    }{%
        \mathbf{z}_k^\dagger \mathbf{M}\mathbf{z}_k
    }
\end{equation}
where $\langle\Psi|\hat{V}|\Psi_k\rangle = \langle\Psi|\hat{\mu}|\Psi_k\rangle$ is the transition dipole moment matrix element and \(\mathbf{v}'\) is the so-called property gradient vector.\cite{Olsen:1985p3235,Sauer:2011} 

In the linear-response formulation of the ODC-12 method (LR-ODC-12),\cite{Copan:2018p4097} the \(\mathbf{E}\) and \(\mathbf{M}\) matrices in \cref{eq:excitation-energy} have the following form:
\begin{equation}
    \label{eq:lr-odc12-hessian-blocks}
    \mathbf{E}
    =
    \begin{pmatrix}
        \mathbf{A}_{11} & \mathbf{A}_{12} & \mathbf{B}_{11} & \mathbf{B}_{12} \\
        \mathbf{A}_{21} & \mathbf{A}_{22} & \mathbf{B}_{21} & \mathbf{B}_{22} \\
        \mathbf{B}_{11}^* & \mathbf{B}_{12}^* & \mathbf{A}_{11}^* & \mathbf{A}_{12}^* \\
        \mathbf{B}_{21}^* & \mathbf{B}_{22}^* & \mathbf{A}_{21}^* & \mathbf{A}_{22}^* \\
    \end{pmatrix}
\end{equation}
\begin{equation}
    \label{eq:lr-odc12-metric-blocks}
    \mathbf{M}
    =
    \begin{pmatrix}
        \mathbf{S}_{11} & \mathbf{0} & \mathbf{0} & \mathbf{0} \\
        \mathbf{0} & \mathbf{1} & \mathbf{0} & \mathbf{0} \\
        \mathbf{0} & \mathbf{0} & -\mathbf{S}_{11}^* & \mathbf{0} \\
        \mathbf{0} & \mathbf{0} & \mathbf{0} & -\mathbf{1} \\
    \end{pmatrix}
\end{equation}
where the $\mathbf{A}_{11}$ and $\mathbf{B}_{11}$ ($\mathbf{A}_{22}$ and $\mathbf{B}_{22}$) matrices contain second derivatives of the DCT energy with respect to parameters of the \(\hat{T}_1\) (\(\hat{T}_2\)) operators, whereas $\mathbf{A}_{12}$ and $\mathbf{B}_{12}$ represent mixed second derivatives. Each block of $\mathbf{E}$ and $\mathbf{M}$ describe coupling between electronic excitations or deexcitations of different ranks. Specifically, the $\mathbf{A}_{11}$, $\mathbf{A}_{12}$, and $\mathbf{A}_{22}$ blocks correspond to interaction of single excitations with single excitations, single excitations with double excitations, and double excitations with each other, respectively. Similarly, the $\mathbf{B}_{11}$ block couples single excitations with single deexcitations, whereas $\mathbf{B}_{12}$ and $\mathbf{B}_{22}$ couple single-double and double-double excitation/deexcitation pairs. Finally, $\mathbf{A}^*_{11}$, $\mathbf{A}^*_{12}$, and $\mathbf{A}^*_{22}$ describe interaction of single and double deexcitations. The solution of the LR-ODC-12 eigenvalue problem \eqref{eq:excitation-energy} has \(\mathcal{O}(O^2V^4)\) computational scaling, where \(O\) and \(V\) are the numbers of occupied and virtual orbitals, respectively.

\subsection{Core-Valence-Separated LR-ODC-12 (\cvs)}
\label{sec:theory:cvs}
The LR-ODC-12 generalized eigenvalue problem \eqref{eq:excitation-energy} can be solved iteratively using one of the multi-root variations of the Davidson algorithm.\cite{Davidson:1975p87,Liu:1978p49} This iterative method proceeds by forming an expansion space for the generalized eigenvectors $\mathbf{z}_k$ starting with an initial (guess) set of unit trial vectors and progressively growing this space until the lowest $N_{\mathrm{root}}$ eigenvectors are converged. While such algorithm is very efficient for computing excitations of electrons in the valence orbitals, it is not suitable for simulations of the X-ray absorption spectra as it would require converging thousands of roots simultaneously to reach the energies necessary to promote core electrons. 

A computationally efficient solution to this problem called core-valence separation (CVS) approximation has been proposed by Schirmer and co-workers within the framework of the ADC methods.\cite{Barth:1985p867,Koppel:1997p4415,Trofimov:2000p483,Plekan:2008p360,Wenzel:2014p1900,Wenzel:2015p214104} In this approach, the occupied orbitals are divided into two sets: {\it core}, corresponding to the orbitals probed by core-level excitations, and {\it valence}, which contain the remaining occupied orbitals. The CVS approximation relies on the energetic and spatial separation of core and valence occupied orbitals\cite{Cederbaum:1980p481,Cederbaum:1980p206} and consists in retaining excitations that involve at least one core orbital while neglecting all excitations from valence orbitals. The CVS approach can be combined with existing implementations of excited-state methods based on the Davidson or Lanczos eigensolvers, providing efficient access to core-level excitation energies. 

We have implemented the CVS approximation within our LR-ODC-12 program. The resulting \cvs algorithm solves the reduced eigenvalue problem of the form
\begin{equation}
    \label{eq:excitation-energy-cvs}
    \mathbf{\tilde{E}}
    \mathbf{z}_k
    =
    \mathbf{\tilde{M}}
    \mathbf{z}_k
    \omega_k
\end{equation}
where $\mathbf{\tilde{E}}$ and $\mathbf{\tilde{M}}$ are the reduced Hessian and metric matrices. These matrices are constructed from the $\mathbf{E}$ and $\mathbf{M}$ matrices in \cref{eq:lr-odc12-hessian-blocks,eq:lr-odc12-metric-blocks} by selecting the matrix elements corresponding to excitations and deexcitations involving at least one core orbital and setting all of the other elements to zero. 
For example, out of all matrix elements of the single excitation block $\mathbf{A}_{11}$ in \cref{eq:lr-odc12-hessian-blocks}, the \cvs Hessian $\mathbf{\tilde{E}}$ includes only the $A_{Ia,Jb}$ matrix elements, where we use the $I, J, K, \ldots$ and $a, b, c, \ldots$ labels to denote core and virtual orbitals, respectively, and reserve $i, j, k, \ldots$ labels for the valence occupied orbitals. Similarly, for the single deexcitation block $\mathbf{A}_{11}^*$, only the $A_{aI,bJ}$ elements are included in the \cvs approximation. 

\begin{figure*}[t!]
   \subfloat[]{\label{fig:cvs_a}\includegraphics[width=0.45\textwidth]{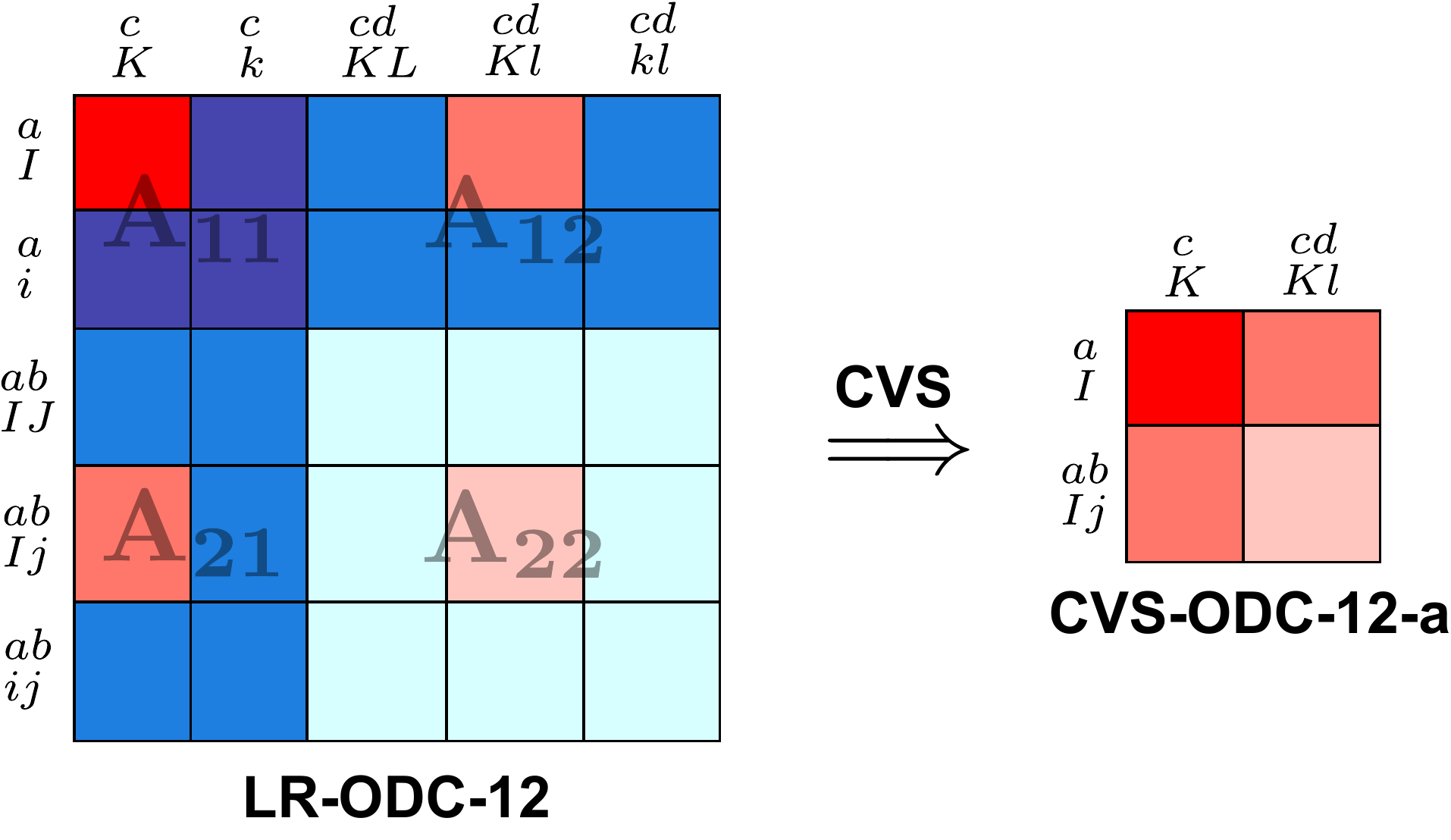}} \qquad
   \subfloat[]{\label{fig:cvs_b}\includegraphics[width=0.45\textwidth]{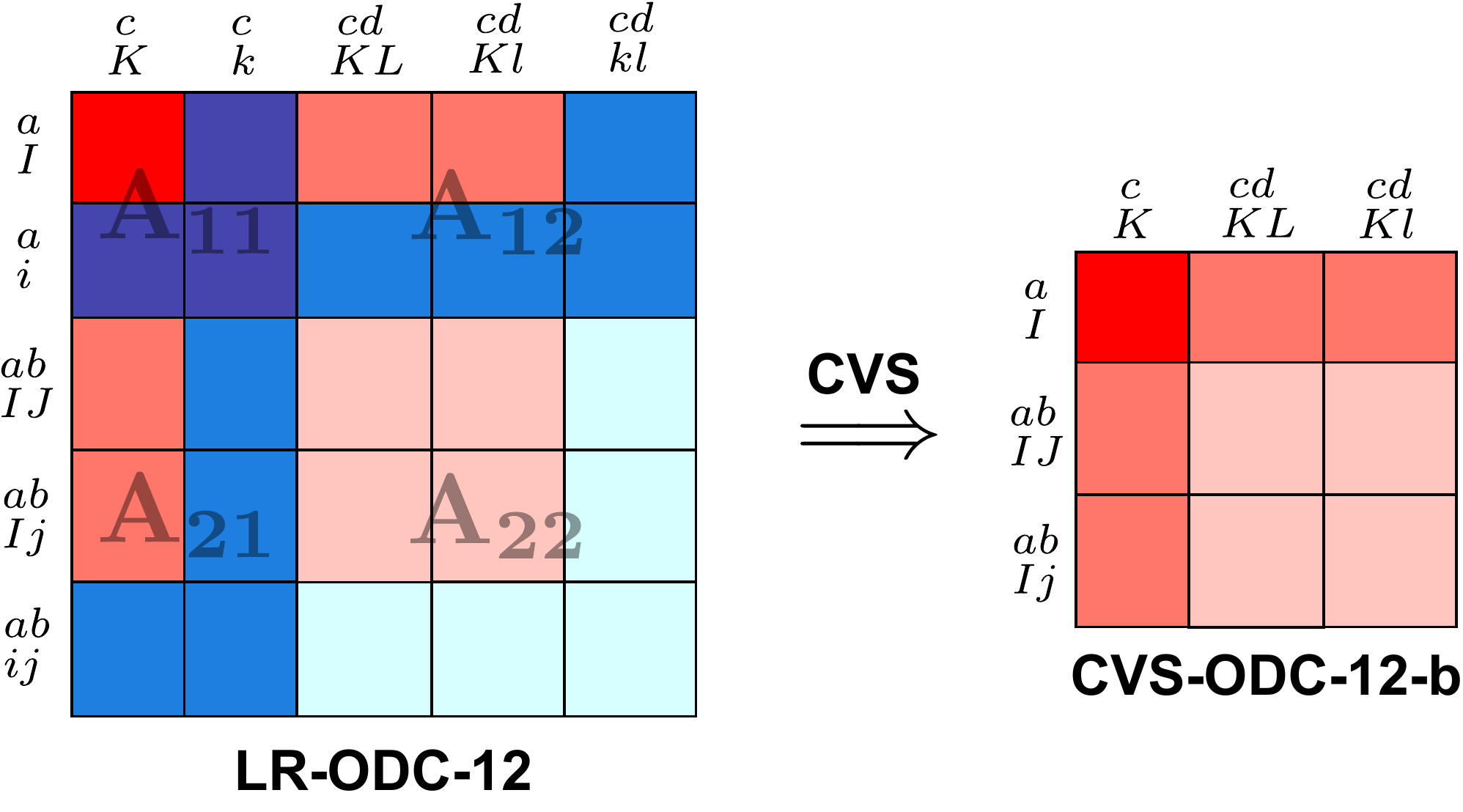}} \quad
   \captionsetup{justification=raggedright,singlelinecheck=false}
   \caption{Illustration of two CVS approximations for the excitation blocks $\mathbf{A_{11}}$, $\mathbf{A_{12}}$, $\mathbf{A_{21}}$, and $\mathbf{A_{22}}$ of the LR-ODC-12 Hessian matrix $\mathbf{E}$ (\cref{eq:lr-odc12-hessian-blocks}). Indices $I, J, K, \ldots$ and $i, j, k, \ldots$ denote core and valence occupied orbitals, while $a, b, c, \ldots$ label virtual orbitals.}
   \label{fig:cvs}
\end{figure*}

For the matrix blocks involving double excitations or deexcitations (e.g., $\mathbf{A}_{22}$ or $\mathbf{A}^*_{22}$), we consider two different CVS schemes. In the first scheme, termed \cvs-a, matrix elements corresponding to double excitations or deexcitations with only one core label are included (e.g., $A_{Ijab,Klab}$ or $A_{Ia,Klab}$), while they are set to zero if a double excitation/deexcitation contains two core indices (e.g., $A_{IJab,KLab}$ or $A_{IJab,Klab}$). This CVS approximation is similar to the one used in the ADC methods.\cite{Barth:1985p867,Koppel:1997p4415,Trofimov:2000p483,Plekan:2008p360,Wenzel:2014p1900,Wenzel:2015p214104} In the second CVS scheme, denoted as \cvs-b, all elements corresponding to double excitations or deexcitations with either one or two core labels are included (e.g., $A_{Ijab,Klab}$, $A_{Ijab,KLab}$, etc.). This CVS approximation has been previously used in combination with the EOM-CCSD method.\cite{Coriani:2015p181103,Lopez:2018} \cref{fig:cvs} illustrates the two CVS approximations for the excitation blocks $\mathbf{A_{11}}$, $\mathbf{A_{12}}$, $\mathbf{A_{21}}$, and $\mathbf{A_{22}}$ of the LR-ODC-12 Hessian matrix in \cref{eq:lr-odc12-hessian-blocks}. Other blocks of the reduced $\mathbf{\tilde{E}}$ and $\mathbf{\tilde{M}}$ matrices can be constructed in a similar way. In \cref{sec:results:accuracy}, we will analyze the accuracy of the \cvs-a and \cvs-b approximations by comparing their results with those obtained from the LR-ODC-12 method.

\section{Computational Details}
\label{sec:comp_details}

The \cvs-a and \cvs-b methods were implemented in a standalone Python program. To obtain the one- and two-electron integrals, our program was interfaced with \textsc{Psi4}\cite{Parrish:2017p3185} and \textsc{Pyscf}.\cite{Sun:2018pe1340} Our implementation of the CVS eigenvalue problem \eqref{eq:excitation-energy-cvs} is based on the multi-root Davidson algorithm,\cite{Davidson:1975p87,Liu:1978p49} where all vectors and matrix-vector products are constructed as outlined in \cref{sec:theory:cvs}. We validated our \cvs implementations against a modified version of the LR-ODC-12 program,\cite{Copan:2018p4097} where the CVS approximation was introduced using the projection technique described by Coriani and Koch.\cite{Coriani:2015p181103}

In all of our computations, all electrons were correlated. For all systems, except ethylene and formic acid, we used the doubly-augmented core-valence d-aug-cc-pCVTZ basis set,\cite{Kendall:1992p6796} where the second set of diffuse functions (d-) was included only for s- and p-orbitals. We refer to this modified basis set as d(s,p)-aug-cc-pCVTZ. In a study of ethylene (\ce{C2H4}), the d(s,p)-aug-cc-pCVTZ basis was used for carbon atoms, while the aug-cc-pVTZ basis was used for the hydrogen atoms. For formic acid (\ce{HCO2H}), we used the d(s,p)-aug-cc-pCVTZ basis for the carbon and oxygen atoms and the aug-cc-pVDZ basis for the hydrogens.

The \cvs X-ray absorption spectra were visualized by plotting the spectral function 
\begin{equation}
    \label{eq:spectral_function}
    T(\omega)
    =
    -\frac{1}{\pi}
    \mathrm{Im}
    \left[
    \sum_{k}
    \frac{|\langle\Psi|\hat{V}|\Psi_k\rangle|^2}{\omega - \omega_k + i\eta}
    \right]
\end{equation}
computed for a range of frequencies $\omega$, where $\omega_k$ are the \cvs excitation energies from \cref{eq:excitation-energy-cvs}, $\eta$ is a small imaginary broadening, and the matrix elements $|\langle\Psi|\hat{V}|\Psi_k\rangle|^2$ are obtained according to \cref{eq:transition-strength}. The simulated spectra were compared to experimental spectra that were digitized using the WebPlotDigitizer program.\cite{WebPlotDigitizer} Electronic transitions were assigned based on the spin and spatial symmetry of excitations, as well as the contributions to the generalized eigenvectors $\mathbf{z}_k$ in \cref{eq:excitation-energy-cvs}.

\section{Results}
\label{sec:results}

\subsection{Accuracy of the CVS Approximations}
\label{sec:results:accuracy}

\begin{table*}[t!]
	\footnotesize
	\caption{Core excitation energies (in eV) and oscillator strengths ($f_{\mathrm{osc}}$) for three lowest singlet (S$_n$) and triplet (T$_n$) excited states computed using LR-ODC-12, as well as its CVS approximations (\cvs-a and \cvs-b). Also shown are mean absolute errors (\mae), standard deviations (\std), and maximum absolute errors (\maxe), relative to LR-ODC-12, computed using data for 12 lowest-energy states (see Supporting Information).}
   \label{tab:lr_vs_cvs}
\begin{tabular}{lcccccc}
        \hline
        \hline
& \multicolumn{2}{c}{LR-ODC-12}  & \multicolumn{2}{c}{\cvs-a} & \multicolumn{2}{c}{\cvs-b}  \\
\cmidrule(lr){2-3}\cmidrule(lr){4-5}\cmidrule(lr){6-7}
Excitation 		& Energy 	& $f_{\mathrm{osc}}\times10^2$ & Energy & $f_{\mathrm{osc}}\times10^2$ & Energy & $f_{\mathrm{osc}}\times10^2$ \\
\hline
\multicolumn{7}{c}{CO (C-edge, STO-3G)} \\
S$_1$ (C$_{1s}\to\pi^*$) 	& 289.13 	& 8.716		& 289.16	& 6.565		& 289.11   	& 8.206		\\
S$_2$        			& 295.55 	& 0.004		& 295.54	& 0.012		& 295.54   	& 0.014		\\
S$_3$        			& 295.84 	& 		& 295.84	& 		& 295.84   	&      		    	\\
T$_1$ (C$_{1s}\to\pi^*$) 	& 288.14 	& 		& 288.16	& 		& 288.14   	&      		    	\\
T$_2$        			& 294.17 	& 		& 294.15	& 		& 294.15   	&      		    	\\
T$_3$        			& 294.77 	& 		& 294.76	& 		& 294.75   	&      		    	\\
\multicolumn{7}{c}{CO (O-edge, STO-3G)} \\
S$_1$ (O$_{1s}\to\pi^*$) 	& 543.43 	& 4.402		& 543.47	& 4.275		& 543.43   	& 4.275		\\
S$_2$        			& 550.03 	& 		& 550.04	& 		& 550.03   	&        	  	\\
S$_3$        			& 553.30 	& 		& 553.31	& 		& 553.30   	&        	  	\\
T$_1$ (O$_{1s}\to\pi^*$) 	& 542.96 	& 		& 542.99	& 		& 542.96   	&      		    	\\
T$_2$        			& 546.94 	& 		& 546.94	& 		& 546.94   	&        	  	\\
T$_3$        			& 549.93 	& 		& 549.93	& 		& 549.93   	&        	  	\\
\multicolumn{7}{c}{\ce{H2O} (6-31G)} \\
S$_1$ (O$_{1s}\to3s$)  		& 541.50 	& 2.452		& 541.64	& 2.383		& 541.49   	& 2.396		\\
S$_2$ (O$_{1s}\to3p$)  		& 543.34 	& 4.945 	& 543.46	& 4.841 	& 543.33   	& 4.862		\\
S$_3$         			& 561.08 	& 10.781	& 561.21	& 10.754	& 561.07   	& 10.770		\\
T$_1$ (O$_{1s}\to3s$)	  	& 540.46 	& 		& 540.54	& 		& 540.45   	&         	 	\\
T$_2$ (O$_{1s}\to3p$)  		& 542.27 	& 		& 542.37	& 		& 542.26   	&          		\\
T$_3$         			& 558.89 	& 		& 558.98	& 		& 558.88   	&          		\\
\mae          			&        	& 		& 0.04		& 0.078		& 0.01 		& 0.031		\\
\maxe         			&        	& 		& 0.13		& 2.151		& 0.02		& 0.510		\\
\std          			&        	& 		& 0.05		& 0.358		& 0.01	 	& 0.090		\\
        \hline
        \hline
\end{tabular}
\end{table*}

We begin by comparing the accuracy of the \cvs-a and \cvs-b approximations described in \cref{sec:theory:cvs}. \cref{tab:lr_vs_cvs} shows core excitation energies and oscillator strengths of CO and \ce{H2O} computed using the full LR-ODC-12 method and the two CVS methods with small basis sets (STO-3G and 6-31G, respectively), for which it was possible to compute the LR-ODC-12 core-excited states using a conventional Davidson algorithm. Out of the two CVS approximations, the best agreement with LR-ODC-12 is shown by \cvs-b that neglects excitations from valence orbitals while retaining all excitations with at least one core label. The superior performance of \cvs-b is reflected by its mean absolute error (\mae) and standard deviation of errors (\std) that do not exceed 0.01 eV, relative to LR-ODC-12, for a combined set of 36 electronic transitions (see Supporting Information for a complete set of data). The \cvs-a approximation, which additionally neglects double excitations from core to virtual orbitals (\cref{fig:cvs}), exhibits much larger \mae and \std values of 0.04 and 0.05 eV, respectively. Although for CO with the STO-3G basis set the \cvs-a errors are in the range of 0.01-0.04 eV, they increase up to 0.13 eV for \ce{H2O}, where a larger 6-31G basis set was used. These errors continue to grow with the size of the one-electron basis set. For example, for transition from the carbon $1s$ orbital to the $\pi^*$ molecular orbital of CO (C$_{1s}\to\pi^*$), the \cvs-a and \cvs-b excitation energies computed using the aug-cc-pVTZ basis set are 289.1 and 288.1 eV, respectively, indicating a large ($\sim$ 1.0 eV) error of the \cvs-a approximation, relative to \cvs-b. Similar results are observed when analyzing performance of the CVS approximations for oscillator strengths, where \cvs-b shows much smaller errors compared to \cvs-a. 

Overall, our results demonstrate that using the \cvs-b approximation has a very small effect on the core-level excitation excitation energies and oscillator strengths, while the errors of the \cvs-a approximation are substantial, especially when large basis sets are used. The small errors introduced by the CVS approximation in \cvs-b are consistent with the CVS errors in the CVS-EOM-CCSD method developed by Coriani and Koch,\cite{Coriani:2015p181103} which employs the same single and double excitation space when constructing the reduced eigenvalue problem. Since the \cvs-a and \cvs-b approximations only differ in the treatment of double excitations from core to valence orbitals and the number of those excitations is usually small, the computational cost of the \cvs-b approximation is similar to \cvs-a. For this reason, we will use the \cvs-b approximation for our study of core-level excitation energies in \cref{sec:results:small_molecules} and will refer to it as \cvs henceforth. 

\subsection{X-Ray Absorption of Small Molecules}
\label{sec:results:small_molecules}

\subsubsection{Excitations Energies}

\begin{table*}[t!]
       \footnotesize
       \caption{Core excitation energies (in eV) and oscillator strengths ($f_{\mathrm{osc}}$) for selected K-edge transitions computed using \cvs. Also shown are best available results from other theoretical methods and experiment. Experimental results are from Refs.\@ \citenum{King:1977p50,Tronc:1980p999,Tronc:1979p137,Hitchcock:1980p1,Puttner:1999p3415,Remmers:1992p3935,Chen:1989p6737,Francis:1994p879,Adachi:1999p427,Domke:1990p122,Schirmer:1993p1136,Hitchcock:1981p4399,Ma:1991p1848,Hitchcock:1979p201,McLaren:1987p1683}. }
   \label{tab:small_molecules}
\begin{threeparttable}
\begin{tabular}{llccccc}
        \hline
        \hline
& & \multicolumn{2}{c}{\cvs}  & \multicolumn{2}{c}{Theory (reference)} & Experiment \\
\cmidrule(lr){3-4}\cmidrule(lr){5-6}\cmidrule(lr){7-7}
Molecule 		& Excitation 		& Energy 	& $f_{\mathrm{osc}}\times10^2$ & Energy & $f_{\mathrm{osc}}\times10^2$ & Energy \\
\hline
\ce{CH4} 		& C$_{1s}\to 3s$	& 290.6 	& 2.19	& 288.2\tnote{a}	& 			& 287.1   		\\
			& C$_{1s}\to 3p$	& 291.4 	& 0.51	& 289.6\tnote{a}	& 			& 288.0   		\\
\ce{C2H4}		& C$_{1s}\to\pi^*$	& 287.2 	& 10.50	& 285.1\tnote{b}	& 9.86\tnote{b}		& 284.7   		\\
			& C$_{1s}\to 3s$	& 290.1 	& 1.27	& 287.7\tnote{b}	& 0.91\tnote{b}		& 287.2   		\\
			& C$_{1s}\to 3p$	& 290.7 	& 4.08	& 288.4\tnote{b}	& 2.75\tnote{b}		& 287.9   		\\
\ce{C2H2}       	& C$_{1s}\to\pi^*$	& 288.2 	& 9.89	& 287.1\tnote{a}	& 			& 285.8   		\\
			& C$_{1s}\to 3s$	& 290.6 	& 1.44	& 289.5\tnote{a}	& 			& 287.9   		\\
			& C$_{1s}\to 3p$	& 291.8 	& 0.70	& 289.8\tnote{a}	& 			& 288.8   		\\
HCN     		& C$_{1s}\to\pi^*$	& 288.5 	& 3.80	& 287.0\tnote{c}	& 			& 286.4   		\\
			& C$_{1s}\to 3s$	& 291.3 	& 1.41	& 289.9\tnote{c}	& 			& 289.1   		\\
			& C$_{1s}\to 3p$	& 293.3 	& 1.07	& 			&			& 290.6   		\\
\ce{H2CO}       	& C$_{1s}\to\pi^*$	& 287.7 	& 6.63	& 286.8\tnote{a}	&			& 285.6   		\\
			& C$_{1s}\to 3s$	& 292.6 	& 1.04	& 291.9\tnote{a}	&			& 290.2   		\\
			& C$_{1s}\to 3p$	& 293.6 	& 3.04	& 292.3\tnote{a}	&			& 291.3   		\\
CO 			& C$_{1s}\to\pi^*$	& 288.6 	& 7.14	& 288.0\tnote{d}	& 16.56\tnote{d}	& 287.3   		\\
			& C$_{1s}\to 3s$	& 294.7 	& 0.45	& 293.0\tnote{b}	& 0.38\tnote{b}		& 292.5   		\\
			& C$_{1s}\to 3p$	& 295.7 	& 0.43	& 294.0\tnote{b}	& 0.95\tnote{b}		& 293.4   		\\
\ce{H2O}		& O$_{1s}\to 3s$	& 538.4 	& 1.61	& 534.5\tnote{d}	& 1.28\tnote{d}		& 534.0   		\\
			& O$_{1s}\to 3p$	& 540.1 	& 3.55	& 536.4\tnote{d}	& 2.62\tnote{d}		& 535.9   		\\
\ce{H2CO}       	& O$_{1s}\to\pi^*$	& 535.2 	& 5.28	& 532.3\tnote{a}	&			& 530.8   		\\
			& O$_{1s}\to 3s$	& 541.2 	& 0.04	& 536.7\tnote{a}	&			& 535.4   		\\
			& O$_{1s}\to 3p$	& 542.0 	& 0.18	& 537.9\tnote{a}	&			& 536.3   		\\
CO			& O$_{1s}\to\pi^*$ 	& 538.5 	& 4.41	& 534.5\tnote{d}	& 8.13\tnote{d}		& 534.1   		\\
			& O$_{1s}\to 3s$	& 543.6 	& 0.14	& 539.7\tnote{b}	& 0.10\tnote{b}		& 538.8   		\\
			& O$_{1s}\to 3p$	& 544.9 	& 0.08	& 540.8\tnote{b}	& 0.10\tnote{b}		& 539.8   		\\
\ce{NH3} 		& N$_{1s}\to 3s$	& 404.2 	& 0.76	& 401.2\tnote{b}	& 0.63\tnote{b}		& 400.8   		\\
			& N$_{1s}\to 3p$	& 405.9 	& 2.96	& 402.9\tnote{b}	& 4.00\tnote{b}		& 402.5   		\\
			& N$_{1s}\to 3p$	& 406.7 	& 0.94	& 403.5\tnote{b}	& 0.63\tnote{b}		& 403.0   		\\
HCN			& N$_{1s}\to\pi^*$	& 403.1 	& 4.26	& 400.6\tnote{c}	&			& 399.7   		\\
			& N$_{1s}\to 3s$	& 407.2 	& 0.05	& 			&			& 402.5   		\\
\ce{N2} 		& N$_{1s}\to\pi^*$	& 403.7 	& 12.49	& 402.0\tnote{e}	& 22.84\tnote{e}	& 401.0   		\\
			& N$_{1s}\to 3s$	& 409.4 	& 0.56	& 407.6\tnote{e}	& 0.45\tnote{e}		& 406.1   		\\
			& N$_{1s}\to 3p$	& 410.5 	& 0.48	& 			&			& 407.0   		\\
HF 			& F$_{1s}\to 4\sigma^*$	& 692.2 	& 2.45	& 689.1\tnote{e}	& 2.26\tnote{e}		& 687.4   		\\
			& F$_{1s}\to 3p_\sigma$ & 696.0 	& 0.74	& 692.8\tnote{e}	& 0.59\tnote{e}		& 690.8   		\\
			& F$_{1s}\to 3p_\pi$	& 696.2 	& 0.64	& 693.0\tnote{e}	& 1.09\tnote{e}		& 691.4   		\\
        \hline
        \hline
\end{tabular}
\begin{tablenotes}
\item[a] IH-FSMRCCSD/aug-cc-pCVXZ (X = T, Q) from Ref.\@ \citenum{Dutta:2014p3656}.
\item[b] fc-CVS-EOM-CCSD/aug-cc-pCVTZ+Rydberg from Ref.\@ \citenum{Lopez:2018}.
\item[c] TD-EOM-CCSD/aug-cc-pVTZ from Ref.\@ \citenum{Nascimento:2017p2951}.
\item[d] CCSDR(3)/aug-cc-pCVTZ+Rydberg from Ref.\@ \citenum{Coriani:2012p022507}.
\item[e] CVS-EOM-CCSD/aug-cc-pCVTZ+Rydberg from Ref.\@ \citenum{Coriani:2015p181103}.
\end{tablenotes}
\end{threeparttable}
\end{table*}

\begin{table*}[t!]
\footnotesize
       \caption{Mean absolute errors (\mae), standard deviations (\std), and maximum absolute errors (\maxe) of the \cvs method computed using excitation energies (in eV) from \cref{tab:small_molecules}, relative to experiment. Experimental results are from Refs.\@ \citenum{King:1977p50,Tronc:1980p999,Tronc:1979p137,Hitchcock:1980p1,Puttner:1999p3415,Remmers:1992p3935,Chen:1989p6737,Francis:1994p879,Adachi:1999p427,Domke:1990p122,Schirmer:1993p1136,Hitchcock:1981p4399,Ma:1991p1848,Hitchcock:1979p201,McLaren:1987p1683}. }
   \label{tab:small_molecules_statistics}
\begin{tabular}{lcccccc}
        \hline
        \hline
		 	&  \multicolumn{3}{c}{Excitation energies}	& \multicolumn{3}{c}{Peak separations}	  \\
\cmidrule(lr){2-4}\cmidrule(lr){5-7}
Excitations 	& \mae 	& \maxe 	& \std 	&  \mae 	& \maxe 	& \std \\
\hline
C-edge		&	2.5&		3.5&		0.5&		0.3& 		1.0& 		0.3\\
N-edge		&	3.5&		4.7&		0.5&		0.5& 		1.3& 		0.5\\
O-edge		&	4.8&		5.8&		0.6&		0.5& 		1.4& 		0.6\\
All		&	3.5&		5.8&		1.1&		0.4& 		1.4& 		0.4\\
        \hline
        \hline
\end{tabular}
\end{table*}

In this section, we use the \cvs method to compute X-ray absorption spectra of small molecules. \cref{tab:small_molecules} shows the \cvs results for 36 core-level transitions of 10 molecules. For comparison, we also show best available theoretical results obtained from various formulations of coupled cluster theory,\cite{Dutta:2014p3656,Lopez:2018,Nascimento:2017p2951,Coriani:2012p022507,Coriani:2015p181103} as well as excitation energies measured in the experiment.\cite{King:1977p50,Tronc:1980p999,Tronc:1979p137,Hitchcock:1980p1,Puttner:1999p3415,Remmers:1992p3935,Chen:1989p6737,Francis:1994p879,Adachi:1999p427,Domke:1990p122,Schirmer:1993p1136,Hitchcock:1981p4399,Ma:1991p1848,Hitchcock:1979p201,McLaren:1987p1683} For all electronic transitions, the \cvs method correctly reproduces the order of peaks observed in the experimental spectra with transitions shifted to higher energies. These systematic shifts are exhibited by many electronic structure methods\cite{Stener:2003p115,Besley:2010p12024,Liang:2011p3540,Zhang:2012p194306,Lestrange:2015p2994,Besley:2016p5018,Nascimento:2017p2951,Coriani:2012p022507} and are usually attributed to basis set incompleteness error and incomplete description of dynamic correlation. As shown in \cref{tab:small_molecules_statistics}, the magnitude of the computed shifts, given by the mean absolute errors (\mae) in the \cvs excitation energies relative to experiment, increases with increasing energy of the K-edge transition. In particular, for C-, N-, and O-edge excitations, \mae increase in the following order: 2.5, 3.5, and 4.8 eV, respectively. Although \mae depend on the type of K-edge excitation, the computed standard deviations of errors (\std) do not change significantly with increasing excitation energy and are relatively small ($\sim$ 0.5 eV), indicating the systematic nature of the observed shifts. This is also supported by the percentage of \mae relative to average excitation energy of each edge, which remains relatively constant for C-, N-, and O-edge transitions (0.87 \%, 0.87 \%, and 0.88 \%, respectively).

The \cvs excitation energies are also shifted relative to reference theoretical results from various coupled cluster methods (\cref{tab:small_molecules}). The \mae values for the computed shifts are 1.6, 2.5, and 3.9 eV for C-, N-, and O-edge transitions, indicating that the shifts in the reference coupled cluster excitation energies are less sensitive to the type of the K-edge transition than those of \cvs. These differences may originate from the different treatment of dynamic correlation and orbital relaxation effects between \cvs and the reference methods, as well as differences in the one-electron basis sets. 

\subsubsection{Peak Separations}

\begin{table*}[t!]
\footnotesize
       \caption{Peak separations (in eV) in the K-edge excitation spectra computed using \cvs. Individual excitation energies are shown in \cref{tab:small_molecules}. Also shown are best available results from other theoretical methods and experiment. Experimental results are from Refs.\@ \citenum{King:1977p50,Tronc:1980p999,Tronc:1979p137,Hitchcock:1980p1,Puttner:1999p3415,Remmers:1992p3935,Chen:1989p6737,Francis:1994p879,Adachi:1999p427,Domke:1990p122,Schirmer:1993p1136,Hitchcock:1981p4399,Ma:1991p1848,Hitchcock:1979p201,McLaren:1987p1683}. }
   \label{tab:small_molecules_relative}
\begin{threeparttable}
\begin{tabular}{llccccc}
        \hline
        \hline
Molecule 		& Excitation 			& \cvs 	& Theory (reference) &  Experiment \\
\hline
\ce{CH4}		& C$_{1s} \to (3p - 3s)$					& 0.8 	& 1.5\tnote{a}		& 1.0 		\\
\ce{C2H4}		& C$_{1s} \to (3s - \pi^*)$					& 2.8 	& 2.6\tnote{b}		& 2.6  		\\
			& C$_{1s} \to (3p - 3s)$					& 0.7 	& 0.7\tnote{b}		& 0.6  		\\
\ce{C2H2} 		& C$_{1s} \to (3s - \pi^*)$					& 2.4 	& 2.5\tnote{a}		& 2.1 		\\
        		& C$_{1s} \to (3p - 3s)$					& 1.2 	& 0.3\tnote{a}		& 0.9 		\\
HCN 			& C$_{1s} \to (3s - \pi^*)$					& 2.8 	& 2.9\tnote{c}		& 2.7 		\\
 			& C$_{1s} \to (3p - 3s)$					& 2.0 	& 			& 1.5 		\\
\ce{H2CO}		& C$_{1s} \to (3s - \pi^*)$					& 4.9 	& 5.1\tnote{a}		& 4.6		\\
        		& C$_{1s} \to (3p - 3s)$					& 1.0 	& 0.4\tnote{a}		& 1.1		\\
CO 			& C$_{1s} \to (3s - \pi^*)$					& 6.1 	& 5.7\tnote{b}		& 5.1 		\\
			& C$_{1s} \to (3p - 3s)$					& 1.0 	& 1.0\tnote{b}		& 1.0 		\\
\ce{H2O} 		& O$_{1s} \to (3p - 3s)$					& 1.7 	& 1.8\tnote{d}		& 1.9		\\
\ce{H2CO}		& O$_{1s} \to (3s - \pi^*)$					& 6.0 	& 4.5\tnote{b}		& 4.6 		\\
			& O$_{1s} \to (3p - 3s)$					& 0.8 	& 1.1\tnote{b}		& 0.9 		\\
CO			& O$_{1s} \to (3s - \pi^*)$					& 5.2 	& 4.7\tnote{b}		& 4.7 		\\
			& O$_{1s} \to (3p - 3s)$					& 1.3 	& 1.1\tnote{b}		& 1.0 		\\
\ce{NH3} 		& N$_{1s} \to (3p - 3s)$					& 1.6 	& 1.7\tnote{b}		& 1.7 		\\
			& N$_{1s} \to (3p - 3p)$					& 0.8 	& 0.7\tnote{b}		& 0.5		\\
HCN			& N$_{1s} \to (3s - \pi^*)$					& 4.1 	& 			& 2.8 		\\
\ce{N2}			& N$_{1s} \to (3s - \pi^*)$					& 5.7 	& 5.6\tnote{e}		& 5.2		\\
			& N$_{1s} \to (3p - 3s)$					& 1.1 	& 			& 0.9 		\\
HF 			& F$_{1s}\to (3p_\sigma - 4\sigma^*)$				& 3.8 	& 3.7\tnote{e}		& 3.4  		\\
			& F$_{1s}\to (3p_\pi - 3p_\sigma)$				& 0.2 	& 0.1\tnote{e}		& 0.6 		\\
        \hline
        \hline
\end{tabular}
\begin{tablenotes}
\item[a] IH-FSMRCCSD/aug-cc-pCVXZ (X = T, Q) from Ref.\@ \citenum{Dutta:2014p3656}.
\item[b] fc-CVS-EOM-CCSD/aug-cc-pCVTZ+Rydberg from Ref.\@ \citenum{Lopez:2018}.
\item[c] TD-EOM-CCSD/aug-cc-pVTZ from Ref.\@ \citenum{Nascimento:2017p2951}.
\item[d] CCSDR(3)/aug-cc-pCVTZ+Rydberg from Ref.\@ \citenum{Coriani:2012p022507}.
\item[e] CVS-EOM-CCSD/aug-cc-pCVTZ+Rydberg from Ref.\@ \citenum{Coriani:2015p181103}.
\end{tablenotes}
\end{threeparttable}
\end{table*}

We now discuss the accuracy of \cvs for simulating energy separation between peaks in X-ray absorption spectra. \cref{tab:small_molecules_relative} shows peak separations computed using \cvs and coupled cluster methods along with experimental results. For most of the electronic transitions, the \cvs method shows a good agreement with experiment predicting peak separations within 0.5 eV from experimental values. This is reflected by its \mae of 0.4 eV, relative to experiment (\cref{tab:small_molecules_statistics}). Contrary to excitation energies, the errors in the \cvs peak spacings exhibit little dependence on the type of the K-edge transition, showing a somewhat smaller \mae for C-edge excitations (0.3 eV) compared to that of the N- and O-edge transitions (0.5 eV). The accuracy of the \cvs peak separations is on par with that of the reference coupled cluster methods, which show \mae of 0.3 eV with respect to experiment. The coupled cluster methods show a similar \std (0.3 eV) and a smaller \maxe (0.7 eV), in comparison to 0.4 eV and 1.4 eV of \cvs, respectively. 

\subsubsection{Simulated Spectra}

\begin{figure*}[t!]
	\includegraphics[width=0.465\textwidth]{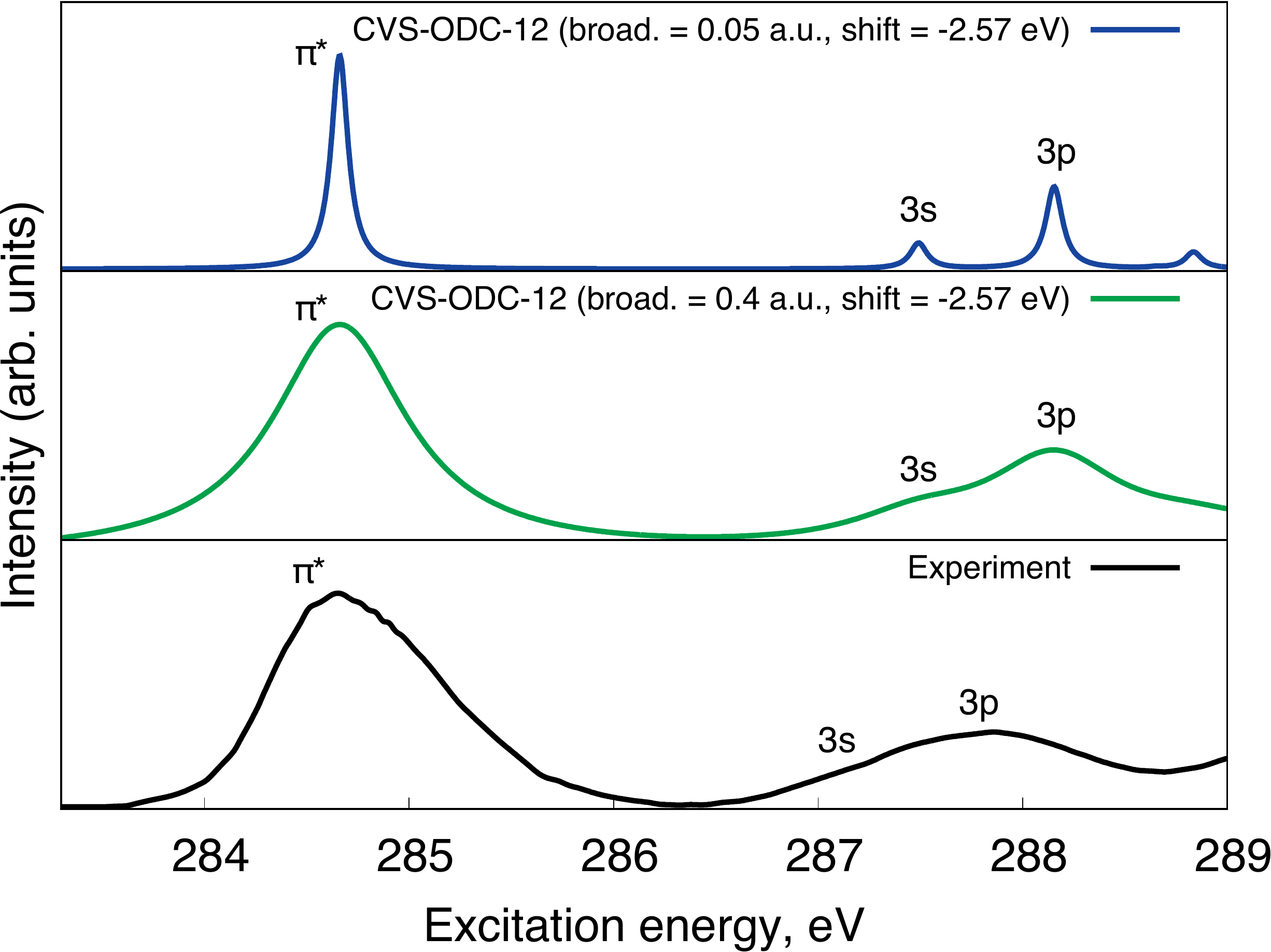}
	\captionsetup{justification=raggedright,singlelinecheck=false}
	\caption{C-edge X-ray absorption spectrum of ethylene computed using \cvs. Results are shown for two broadening parameters (see \cref{sec:comp_details} for details) and are compared to experimental spectrum from Ref.\@ \citenum{McLaren:1987p1683}. The \cvs spectrum was shifted by $-$2.57 eV to reproduce position of the C$_{1s}\to\pi^*$ peak in the experimental spectrum. See \cref{tab:small_molecules} and the Supporting Information for the \cvs excitation energies and oscillator strengths.}
   \label{fig:spectrum_c2h4}
\end{figure*}

In this section, we compare X-ray absorption spectra computed using \cvs with those obtained from experiment for three polyatomic molecules: ethylene (\ce{C2H4}), formaldehyde (\ce{H2CO}), and formic acid (\ce{HCO2H}). Since basis sets employed in our study do not incorporate Rydberg functions, we only consider regions of spectra dominated by core-level transitions into the low-lying $\pi^*$, $3s$, and $3p$ orbitals. In addition to reporting the computed spectra, we provide the spectral data for all three molecules in the Supporting Information. 

\cref{fig:spectrum_c2h4} shows the C-edge spectra of ethylene computed using \cvs for two broadening parameters along with an experimental spectrum from Ref.\@ \citenum{McLaren:1987p1683}. The simulated spectra were shifted by $-$2.57 eV to align the position of the first peak with the one in the experimental spectrum. The shifted \cvs spectra show a good agreement with experiment, reproducing the separation and relative intensity of the C$_{1s}\to\pi^*$, $3s$, and $3p$ transitions. The \cvs method overestimates the relative positions of the $3s$ and $3p$ peaks by $\sim$ 0.2 eV, which is consistent with its \mae of 0.3 eV from \cref{tab:small_molecules_statistics}. 

\begin{figure*}[t!]
   \subfloat[]{\label{fig:spectrum_h2co_c}\includegraphics[width=0.45\textwidth]{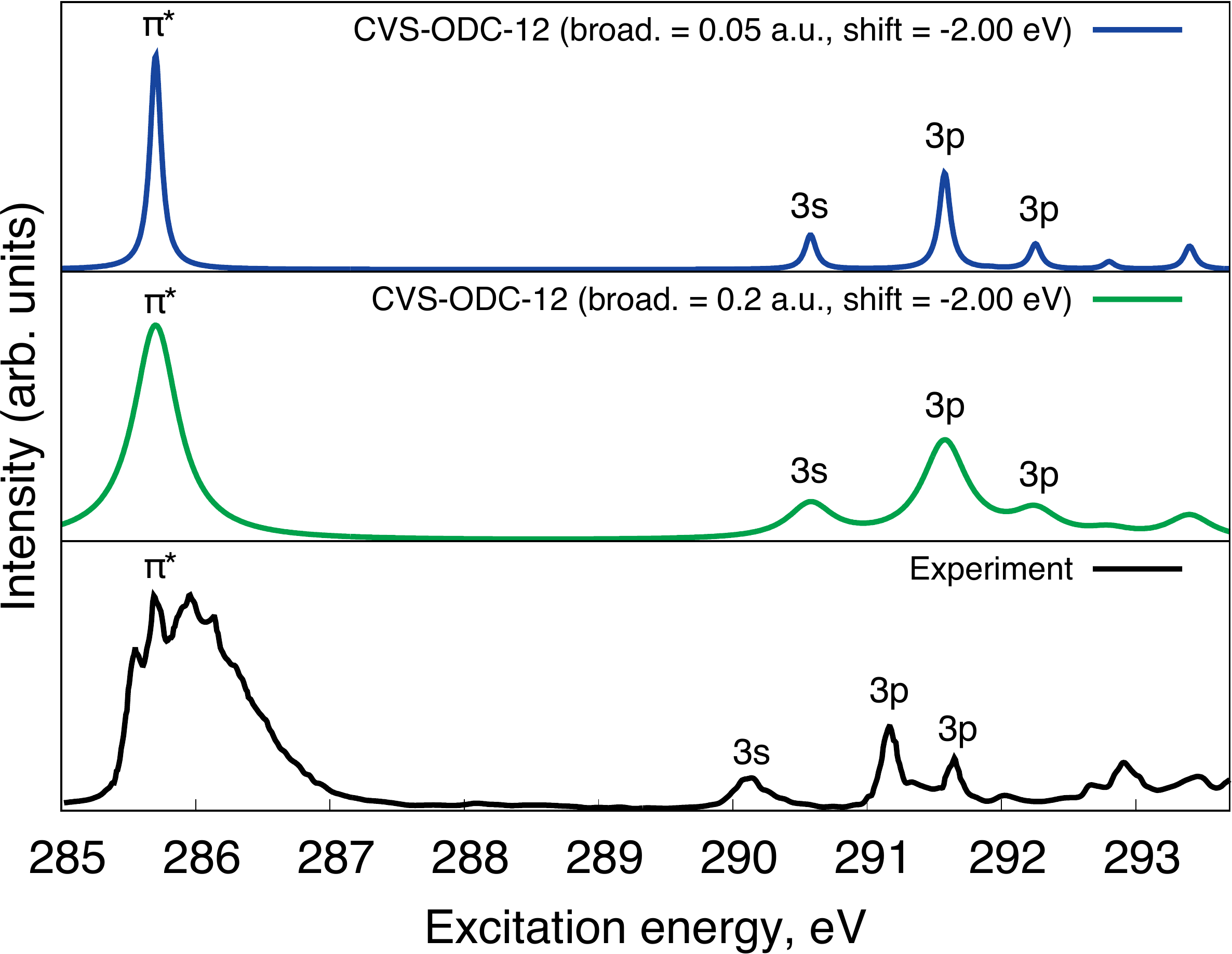}} \qquad
   \subfloat[]{\label{fig:spectrum_h2co_o}\includegraphics[width=0.465\textwidth]{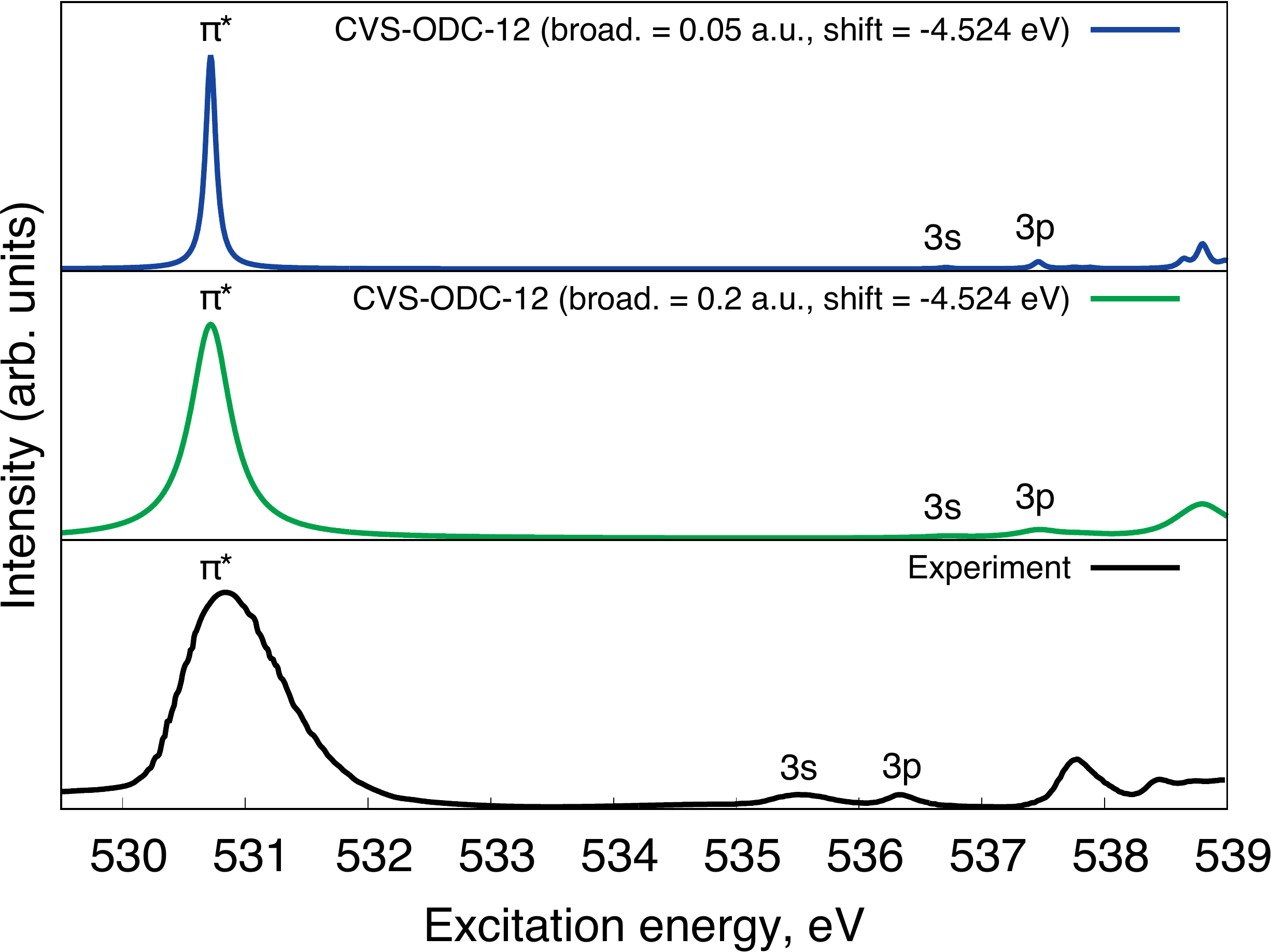}} \quad
   \captionsetup{justification=raggedright,singlelinecheck=false}
	\caption{C-edge (\ref{fig:spectrum_h2co_c}) and O-edge (\ref{fig:spectrum_h2co_o}) X-ray absorption spectra of formaldehyde computed using \cvs. Results are shown for two broadening parameters (see \cref{sec:comp_details} for details) and are compared to experimental spectra from Ref.\@ \citenum{Remmers:1992p3935}. The \cvs spectra were shifted to reproduce positions of the C$_{1s}\to\pi^*$ and O$_{1s}\to\pi^*$ peaks in the experimental spectra. See \cref{tab:small_molecules} and the Supporting Information for the \cvs excitation energies and oscillator strengths.}
   \label{fig:spectrum_h2co}
\end{figure*}

\cref{fig:spectrum_h2co} reports the computed C-edge and O-edge spectra of formaldehyde. Aligning positions of the C$_{1s}\to\pi^*$ and O$_{1s}\to\pi^*$ peaks with those in the experimental spectrum\cite{Remmers:1992p3935} requires shifting the \cvs spectra by $-$2.00 and $-$4.524 eV, respectively, consistent with \mae of 2.5 and 4.8 eV for excitation energies reported in \cref{tab:small_molecules_statistics}. After the shift, the simulated C-edge spectrum reproduces position of the $3s$ and $3p$ peaks within 0.3-0.4 eV from experiment. The relative intensities of these transitions also agree well with those observed in the experimental spectrum. For oxygen edge, the agreement between \cvs and experiment is worse: the relative energies of the $3s$ and $3p$ transitions are overestimated by $\sim$ 1.3 eV. In addition, in the simulated spectrum the relative intensity of the $3s$ peak is significantly lower than the one obtained in the experiment. 
The energy spacing between the $3s$ and $3p$ transitions (0.8 eV) is in a good agreement with that from the experimental spectrum (0.9 eV).

\begin{figure*}[t!]
   \subfloat[]{\label{fig:spectrum_hcooh_c}\includegraphics[width=0.45\textwidth]{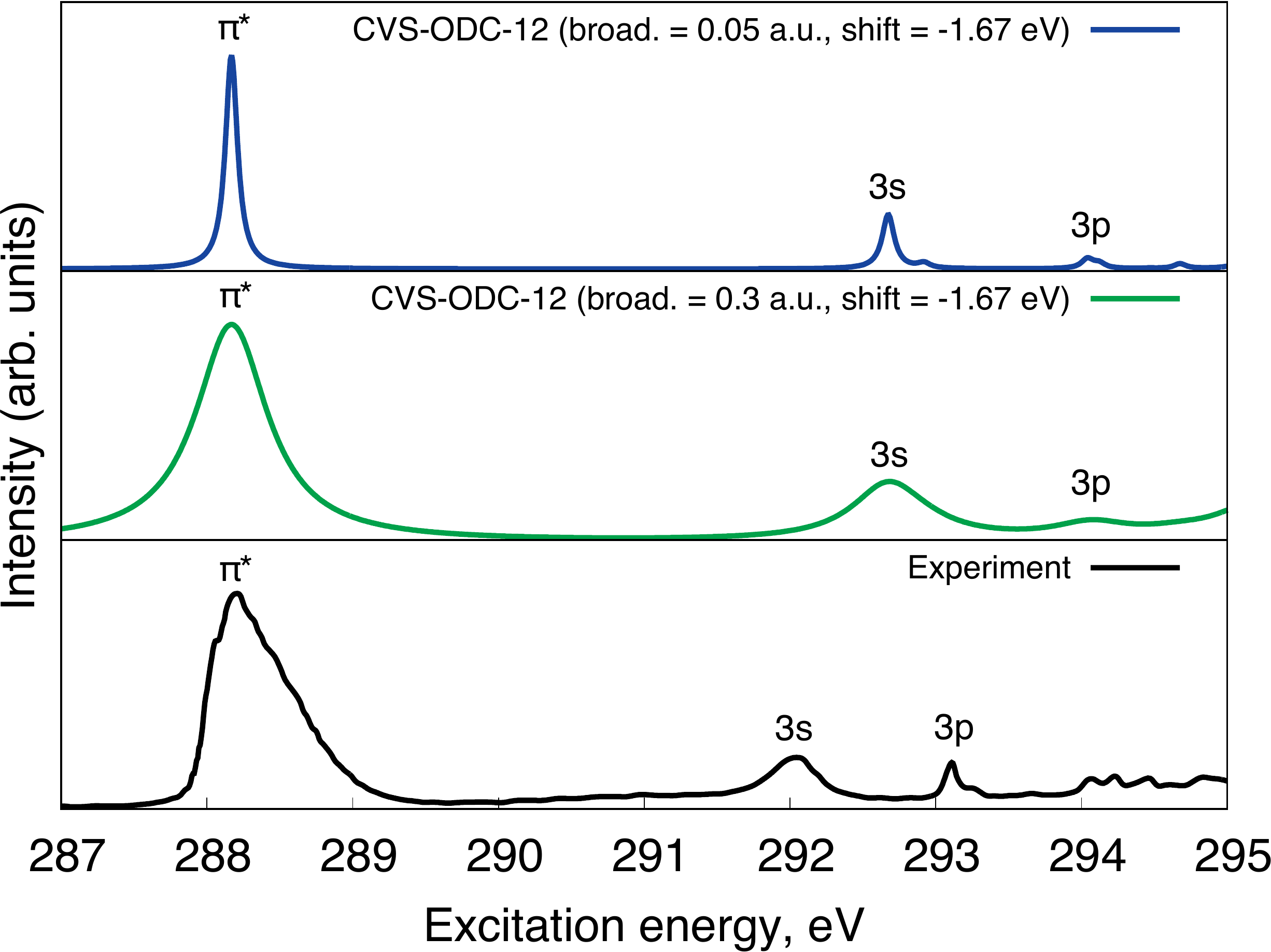}} \qquad
   \subfloat[]{\label{fig:spectrum_hcooh_o}\includegraphics[width=0.45\textwidth]{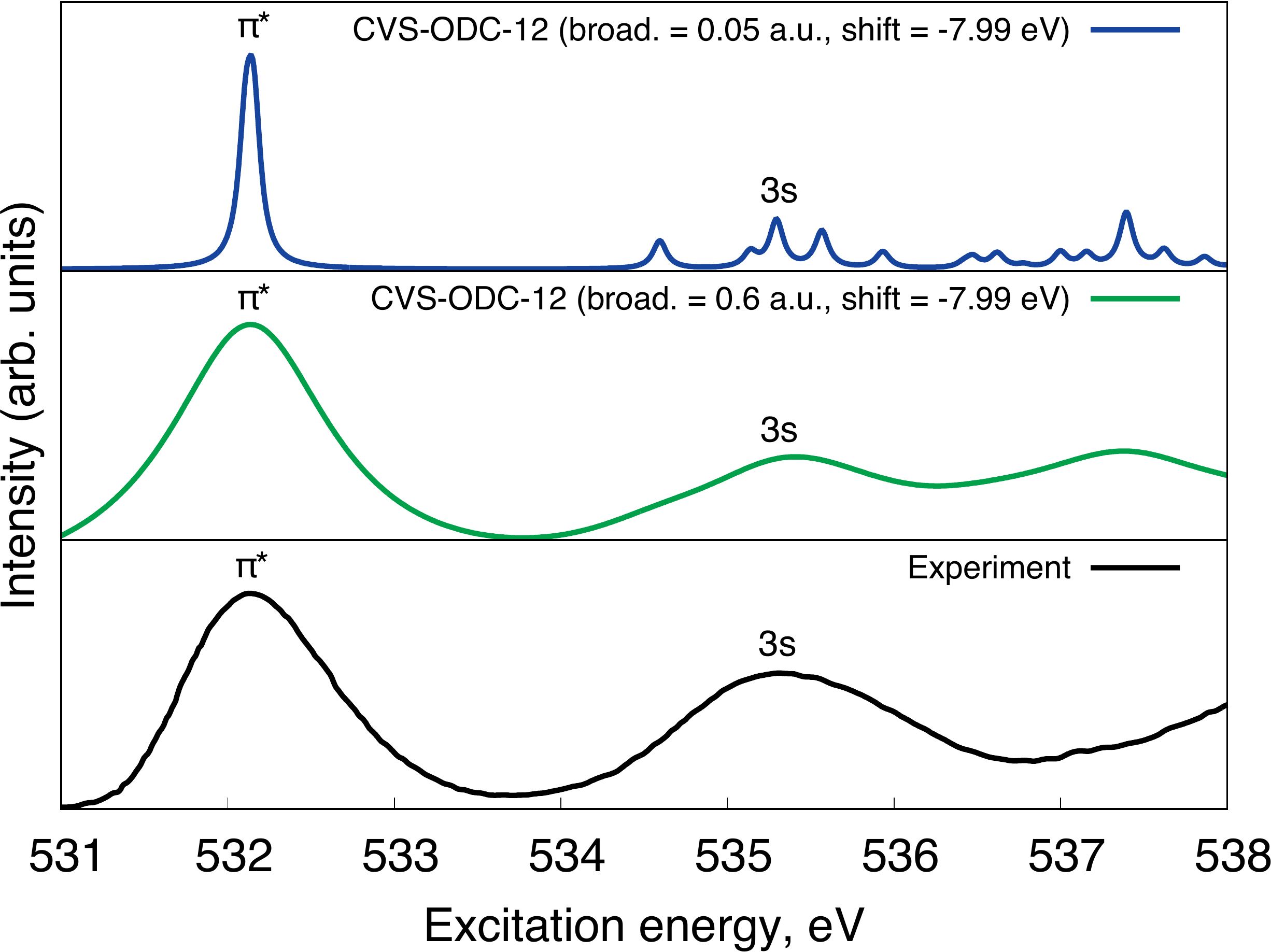}} \quad
   \captionsetup{justification=raggedright,singlelinecheck=false}
	\caption{C-edge (\ref{fig:spectrum_hcooh_c}) and O-edge (\ref{fig:spectrum_hcooh_o}) X-ray absorption spectra of formic acid computed using \cvs. Results are shown for two broadening parameters (see \cref{sec:comp_details} for details) and are compared to experimental spectra from Ref.\@ \citenum{Prince:2002p159}. The \cvs spectra were shifted to reproduce positions of the C$_{1s}\to\pi^*$ and O$_{1s}\to\pi^*$ peaks in the experimental spectra. See the Supporting Information for the \cvs excitation energies and oscillator strengths.}
   \label{fig:spectrum_hcooh}
\end{figure*}

Finally, we consider formic acid as an example of a molecule with a more complicated X-ray absorption spectra. The computed \cvs spectra for carbon and oxygen edge are shown in \cref{fig:spectrum_hcooh}. Aligning simulated and experimental C-edge spectra requires a shift of $-$1.67 eV, while a large shift of $-$7.99 eV is needed for the oxygen edge. For carbon edge, the \cvs method shows a reasonable agreement with experiment for the $3s$ transition with an error of $\sim$ 0.5 eV, while a larger error of $\sim$ 0.8 eV is observed for the third peak ($3p$). When considering the oxygen edge, the experimental spectrum\cite{Prince:2002p159} shows two broad signals at 532.1 and 535.3 eV attributed to the O$_{1s}\to\pi^*$ and O$_{1s}\to3s$ transitions, respectively. The \cvs O-edge spectrum reveals that the second signal originates from several closely spaced peaks corresponding to excitations into $3s$ orbitals of all carbon and oxygen atoms, with significant contributions of excitations to $3p$ orbitals. When using a large broadening parameter, these transitions form a broad signal with a maximum at 535.4 eV, in a very good agreement with experimental spectrum. 

\section{Conclusions}
\label{sec:conclusions}

In this work, we have presented a new approach for simulations of X-ray absorption spectra based on linear-response density cumulant theory (LR-DCT). Our new method combines the LR-ODC-12 formulation of LR-DCT with core-valence separation approximation (CVS) that allows to efficiently access high-energy core-excited states. We considered two CVS approximations of LR-ODC-12 (\cvs) that incorporate different types of excitations from core to virtual orbitals and compared their results with core-level excitation energies obtained from the full LR-ODC-12 method. Our results demonstrated that including double core-virtual excitations is crucial to maintain high accuracy of the CVS approximation for K-edge excitation energies, especially when using large one-electron basis sets.

We have used the \cvs method to compute X-ray absorption spectra of several small molecules and compared them to spectra obtained from experiment. The \cvs method shows a good agreement with experiment for spacings between transitions and their relative intensities, but the computed spectra are systematically shifted to higher energies. The magnitude of these shifts increases with increasing energy of the K-edge transition: for C-, N-, and O-edge transitions the \cvs excitation energies are shifted by $\sim$ 2.5, 3.5, and 4.8 eV, on average. However, the relative distances between transitions depend much less on the K-edge excitation energy, reproducing experimental peak spacings within 0.3-0.5 eV for most of the computed transitions. We have compared the \cvs results with X-ray absorption spectra computed using different formulations of excited-state coupled cluster theory with single and double excitations (CCSD). For peak spacings and intensities, the \cvs and CCSD methods show similar performance, but the CCSD spectra exhibit smaller shifts, particularly for N- and O-edge transitions. An important advantage of \cvs over the CCSD methods is that the former is based on the diagonalization of a Hermitian matrix, which enables efficient computation of transition intensities and guarantees that the resulting excitation energies have real values, provided that the matrix is positive semidefinite. Moreover, the \cvs and CCSD methods have the same $\mathcal{O}(N^6)$ computational scaling with the size of the one-electron basis set $N$.

Overall, our results suggest that \cvs is a useful method for qualitative and semi-quantitative predictions of X-ray absorption spectra of molecules. To improve the quality of the \cvs excitation energies further, we plan to combine this method with frozen-core approximation, as suggested in the recent work by Lopez \mbox{et al.}\cite{Lopez:2018} within the framework of CVS-approximated EOM-CCSD. We also plan to develop an efficient implementation of \cvs that incorporates treatment of relativistic effects and benchmark its performance for open-shell molecules and transition metal compounds. 

\acknowledgement
This work was supported by start-up funds provided by the Ohio State University. 
 
\suppinfo

Table comparing the \cvs approximations with the full LR-ODC-12 method and tables with spectral data for ethylene, formaldehyde, and formic acid.

\providecommand{\latin}[1]{#1}
\makeatletter
\providecommand{\doi}
  {\begingroup\let\do\@makeother\dospecials
  \catcode`\{=1 \catcode`\}=2 \doi@aux}
\providecommand{\doi@aux}[1]{\endgroup\texttt{#1}}
\makeatother
\providecommand*\mcitethebibliography{\thebibliography}
\csname @ifundefined\endcsname{endmcitethebibliography}
  {\let\endmcitethebibliography\endthebibliography}{}

\end{document}